# Observation of the transition from lasing driven by a bosonic to a fermionic reservoir in a GaAs quantum well microcavity

S. Brodbeck,[1] H. Suchomel,[1] M. Amthor,[1] T. Steinl,[1] M. Kamp,[1] C. Schneider,[1] and S. Höfling[1,2]

[1]*Technische Physik, Physikalisches Institut and Wilhelm Conrad Röntgen-Research Center for Complex Material Systems, Universität Würzburg, Am Hubland, D-97074 Würzburg, Germany*

[2]*SUPA, School of Physics and Astronomy, University of St. Andrews, St. Andrews, KY 16 9SS, United Kingdom*

We show that by monitoring the free carrier reservoir in a GaAs-based quantum well microcavity under non-resonant pulsed optical pumping, lasing supported by a fermionic reservoir (photon lasing) can be distinguished from lasing supported by a reservoir of bosons (polariton lasing). Carrier densities are probed by measuring the photocurrent between lateral contacts deposited directly on the quantum wells of a microcavity that are partially exposed by wet chemical etching. We identify two clear thresholds in the input-output characteristic of the photoluminescence signal which can be attributed to polariton and photon lasing, respectively. The power dependence of the probed photocurrent shows a distinct kink at the threshold power for photon lasing due to increased radiative recombination of free carriers as stimulated emission into the cavity mode sets in. At the polariton lasing threshold on the other hand, the nonlinear increase of the luminescence is caused by stimulated scattering of exciton-polaritons to the ground state which do not contribute directly to the photocurrent.

Quantum well microcavities[1] have emerged as systems for the study of lasing transitions triggered by stimulated scattering rather than stimulated emission. Strong coupling of quantum well excitons and cavity photons leads to the formation of mixed light-matter quasiparticles, the so-called exciton-polaritons (polaritons). At sufficiently high densities, stimulated scattering from a bosonic exciton reservoir results in polariton lasing from the ground state at $k_\parallel = 0$[2,3]. Spectral signatures of this transition include a nonlinear increase in emission from the ground state, linewidth narrowing, and a persistent blueshift above threshold due to repulsive polariton-polariton and polariton-reservoir interactions[3]. At high excitation powers, a second nonlinearity may be observed in the same sample. This is commonly attributed to photon lasing in the weak coupling regime as strong coupling is bleached[4-6]. Consequently, the two lasing transitions would be caused by different physical processes and supported by different reservoirs since photon lasing is due to stimulated emission of photons into the cavity mode. In this case, free electrons and holes would act as a reservoir with fermionic nature. The different nature of the reservoirs for polariton and photon lasing explains the occasionally used terminology of bosonic and fermionic lasing, respectively[7].

There are other possible explanations for observing two thresholds in a microcavity, like spin-polarized[8] or multi-mode lasing[9]. More recently, a BCS-type transition was proposed as explanation for the second treshold[10]. Accessing excitons and free carriers should reveal whether the two lasing transitions are indeed supported by different reservoirs since the carrier density is expected to be clamped at threshold for photon lasing[11-14] and the exciton density is at least partially clamped for polariton lasing[15,16]. Measurements of carrier or exciton densities can additionally give insight into relaxation rates and into coupling between the two reservoirs, for instance by modelling with coupled rate equations[13,17,18]. So far, reservoir densities have only been examined separately for photon and polariton lasing transitions for example by measuring the spontaneous emission in photon lasers to monitor the carrier density[12-14]. For polariton lasers, exciton and free carrier densities have been deduced from the transmission of a THz pulse[16], but this technique has not yet been applied to investigating the second threshold.

Here, we probe free carrier densities across the two thresholds in a quantum well microcavity via the in-plane photocurrent perpendicular to the direction of optical excitation and sample luminescence. Electrical contacts are deposited on the sidewalls of an etched mesa to directly contact the quantum wells. The photocurrent is proportional to the carrier density assuming a constant mobility. Excitons and exciton-polaritons are charge neutral and therefore do not contribute to the photocurrent.

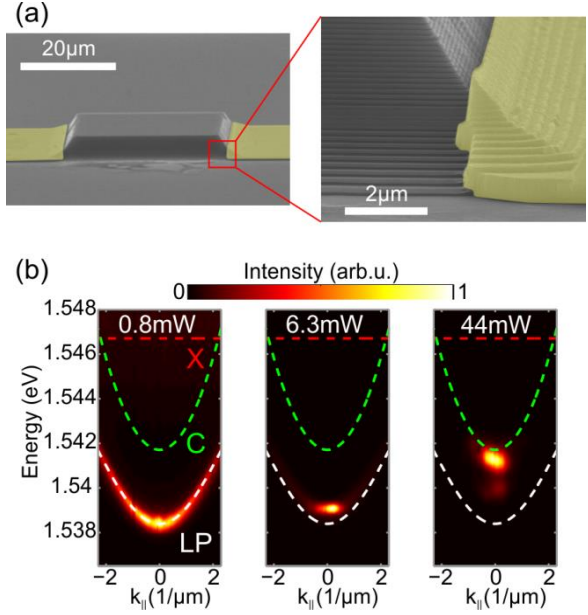

Fig. 1: (a) Scanning electron microscope images of a mesa with electrical contacts (yellow overlay) on the sidewalls. Left image shows the whole device and the right image a zoom-in of the contact to the quantum wells. (b) Momentum resolved PL spectra at three different excitation powers representing the linear polariton (left), polariton laser (middle) and photon laser regimes (right). White dashed lines are theoretical dispersions of the lower polariton fitted to the low power spectrum. Green (red) dashed lines are bare photon (exciton) dispersions used in the fit.

We study an AlAs/Al$_{0.2}$Ga$_{0.8}$As microcavity incorporating 12 GaAs quantum wells with 13nm width embedded in the three central antinodes of the light field. The quantum wells are separated by 4nm wide AlAs barriers and the cavity is formed by a $\lambda/2$–wide AlAs layer. All layers are nominally undoped except for the GaAs substrate which is n-doped. A Rabi splitting of 9.5meV and Q-factor of 10 000 were experimentally determined for this sample[19]. A detuning between bare photon and exciton energies of $E_C - E_X = -5.0$meV was chosen for all measurements. Square mesas with edge length d=50µm were defined by electron beam lithography and wet chemical etching. Electrical contacts (Au-Ti-Pt alloy) were deposited on two opposite sides of the mesa onto the sloped sidewalls. This results in direct Schottky-contacts to the GaAs quantum wells which are partially exposed by the wet etching. The etching was stopped well above the n-doped GaAs substrate to avoid its contribution to the electrical signal. Scanning electron microscope images of a contacted mesa are shown in Fig. 1(a).

Measurements were performed on a Fourier space micro PL setup providing energy- as well as momentum-resolution. The sample temperature was kept at 5K during the measurements. A Ti:Sapphire laser with a spot size of approx. 50µm at a repetition rate of 82MHz and with a pulse length of about 40ps was used for excitation. We did not observe two thresholds under continuous wave (cw) excitation due to limited cw-excitation powers[20]. Our pulse length significantly exceeds the lifetimes of cavity photons (≈4ps) and ground state polaritons (≈5.5ps) and it is comparable to previously reported signal delay times after arrival of the pump pulse for polariton and photon lasers[6,16,21]. Therefore, the system should be not too far from equilibrium during an excitation pulse. The laser energy was tuned to the first reflectance minimum outside the stopband at 1.63eV which is far below the absorption edge of $Al_{0.2}Ga_{0.8}As$ or AlAs. Consequently, carriers are only excited in the GaAs quantum wells and substrate. Due to the low sample temperature, intrinsic carrier densities are negligible. The detector of the setup was calibrated by tuning the excitation laser to the emission wavelength of the microcavity at 1.54eV and reflecting it off a mirror with known reflectivity mounted in the cryostat used for measurements on the microcavity sample. Measuring the total number of counts on the charge coupled device of the detector for the reflected laser signal as well as the power of the incident laser before the objective yields the factor for conversion from counts per second to power emitted from the measured sample. This power can then be converted to emitted photons per pulse given the energy of the emission and the repetition rate of the excitation laser. The calibration is done at normal incidence since the PL spectra are quantitavely analyzed at $k_\parallel = 0$ which corresponds to 0° emission angle.

A voltage source was connected to the contacts on the sidewalls of the investigated mesa and the voltage drop across a series resistance with $R = 5.6\text{k}\Omega$ was recorded to determine the current between the contacts. There was no measurable current without illumination by the excitation laser. Due to the long reservoir lifetimes and short photon and polariton lifetimes, the current is integrated over pulses of hundreds of ps length while the luminescence after the lasing transitions is integrated over pulses of only few tens of ps length. Nevertheless, photon lasing should have a notable impact on the integrated carrier density as we have confirmed by solving the standard rate equations for a photon laser for pulsed excitation[11,18].

We observe two thresholds in the PL signal when increasing the excitation power. Fig. 1(b) shows momentum resolved PL spectra for the three different regimes below the first threshold (left spectrum), slightly above the first threshold (middle spectrum), and above the second threshold (right spectrum) when no voltage source is connected to the contacts (open circuit configuration). We attribute the first threshold to the onset of polariton lasing and the second threshold to photon lasing. Emission from both polariton and photon lasers is observed around the second threshold due to the pulsed excitation scheme[21]. In previous time-resolved experiments, emission from the photon laser was shown to set in first after the excitation pulse with emission from the polariton laser following later[6].

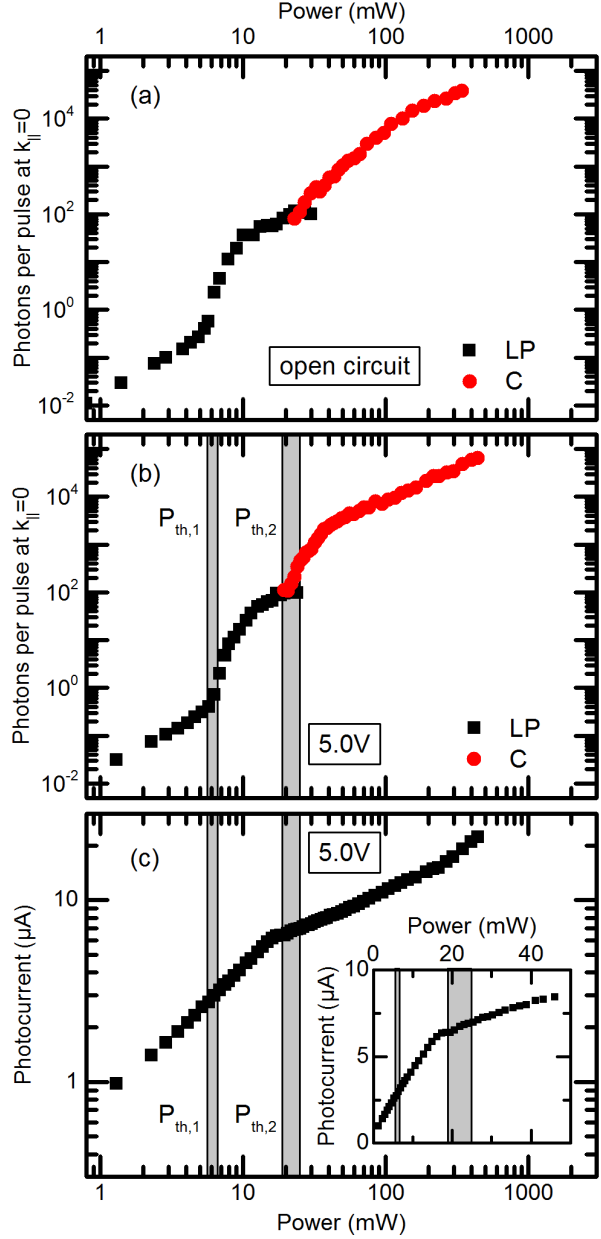

Fig. 2: (a) Average number of photons emitted per pulse as a function of excitation power without external bias. At 20mW to 30mW, emission from both the polariton (black squares) and the photon laser (red circles) is visible due to the pulsed excitation scheme. (b) Same as (a) for an external bias of 5V corresponding to an in-plane electric field of 1kV/cm. (c) Lateral photocurrent as a function of excitation power for an external bias of 5V showing a clear kink at the second threshold. Inset shows the photocurrent for P<50mW in linear scale.

The input-output characteristic for the open circuit configuration is plotted in Fig. 2(a). For this analysis, spectra at $k_\parallel = (0 \pm 0.1)\mu m^{-1}$ are extracted from the momentum resolved PL spectra and fitted with one or two Lorentzian peaks. The peak that is lower in energy is attributed to emission from the lower polariton and the peak at higher energy appearing at around 20mW[18] to emission from the cavity mode, see also Fig. 3(c). The second peak is slightly redshifted compared to the bare cavity energy deduced from a coupled oscillator fit to low power spectra, cf. Fig. 1(b), due to the modified refractive index of the quantum well material at higher excitation powers[5]. Two clear thresholds can be seen in Fig. 2(a) at $(5.2 \pm 0.5)$mW and $(24.9 \pm 3.0)$mW. We define the second threshold at the excitation power for which the number of photons emitted from the cavity mode exceeds that of photons emitted from the lower polariton ground state and the margin of error is given by the range of excitation powers for which both peaks are comparably bright. The second threshold as defined here separates the low power regime dominated by polaritons from the high power regime dominated by emission from the cavity mode. We note that the most appropriate threshold condition for photon lasing would be the power where there is one photon in the cavity mode, but emission from the photon laser at threshold is outshone by emission from the polariton laser. The average number of polaritons per pulse at $k_\parallel = 0$ is on the order of 1 at the first threshold[3] and the number of photons emitted from the cavity mode at the second threshold is about 100.

Next, we apply a constant bias $U = 5$V between the two contacts on the investigated mesa which corresponds to an electric field $F \approx 1$kV/cm in the plane of the quantum wells. Fig. 2(b) depicts the resulting input-output characteristic with two clear thresholds at $(6.1 \pm 0.5)$mW and $(21.9 \pm 3.0)$mW. The average numbers of emitted photons at the two threshold powers are again on the order of 1 and 100, respectively. The increase of the polariton lasing threshold can be reproduced in a rate equation model which includes field ionization losses[18,22]. For $U = 5$V, the input-output curve of emission attributed to the cavity mode can be fit with a rate equation model that is commonly used for photon lasers[23] with the fraction of spontaneous emission into the cavity mode $\beta = 0.11\%$ as fitting parameter. Extrapolating the fit to lower excitation powers suggests a photon lasing threshold around 14mW, but the fit has a large uncertainty as only data at high excitation powers are available. In Fig. 2(a), the intensity emitted from the cavity mode rises less abruptly and the input-output curve cannot be fitted for a reasonable range of parameters.

Fig. 2(c) shows the photocurrent measured between the lateral contacts for $U = 5$V as a function of excitation power. There is a clear kink at around 19mW with approximately linear dependencies for lower and higher excitation powers. A linear least squares fit to 15 data points in the intervals from 1mW to 11mW and from 22mW to 43mW yields slopes of $(0.377 \pm 0.004)$μA/mW and $(0.075 \pm 0.002)$μA/mW, respectively. The density of free carriers $n_{eh}$ can be generally described, i.e. in strong as well as weak coupling regimes, by a rate equation $dn_{eh}/dt = P - \gamma n_{eh}$[11,24,25] where $P$ is the pump rate and $\gamma$ the loss rate. In the steady state ($dn_{eh}/dt = 0$), this yields $n_{eh} = P/\gamma$. The losses are composed of spontaneous, stimulated and non-radiative recombination as well as exciton formation. The kink in the photocurrent and reduced slope for large excitation powers indicate a sudden increase of one of the loss channels. Since the kink coincides with the emergence of bright emission close to the bare cavity mode, this is likely due to the onset of stimulated emission which is consistent with the common interpretation that the second threshold is caused by photon lasing supported by a reservoir of free carriers. We do not observe full clamping of the carrier density at the photon laser threshold. Solving the rate equations for a standard photon laser[11] for excitation with 40ps long pulses suggests that carrier densities at the beginning of the pulse can exceed the transparency density because stimulated emission only sets in once the photon mode is sufficiently populated by spontaneous emission[18]. At the onset of photon lasing, which occurs before the end of the excitation pulse, the carrier density is pinned to the transparency density. Therefore, the total integrated carrier density slightly increases with increasing power even after the lasing threshold for pulsed excitation. Additionally, the carrier reservoir may not be fully thermalized due to the non-resonant excitation which also leads to incomplete pinning of the carrier density[13,17].

At the polariton lasing threshold, no impact on the photocurrent is observed which implies that the carrier reservoir is neither affected by the polariton density in the ground state nor by the density of cold excitons in the reservoir. This means that the exciton and carrier reservoirs are necessarily out of equilibrium. In principle, applying an in-plane electric field introduces ionization losses for excitons[22] which feed the free carrier reservoir. This can allow for equilibration between the two reservoirs. We model our system by a set of rate equations which account for field ionization in the exciton reservoir. By including a hot and a cold exciton reservoir in our formalism, we find that stimulated scattering to the ground state has a negligible impact on the carrier reservoir density even for notable exciton ionization rates[18].

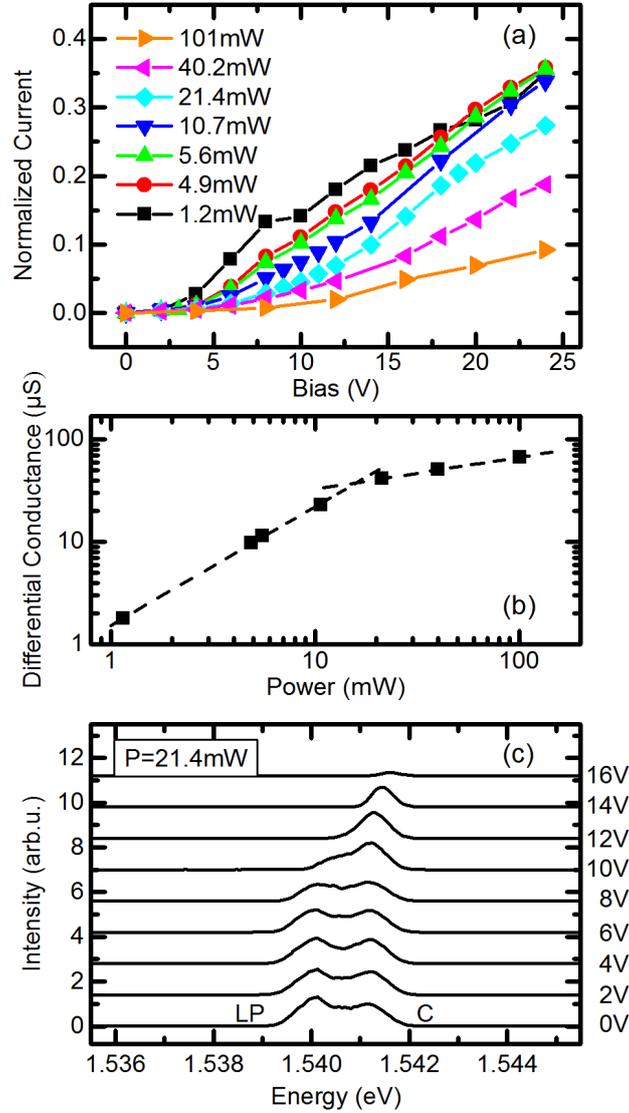

Fig. 3: (a) Current-voltage characteristics at different excitation powers. The current is normalized to the total number of excited carriers at each power. (b) Differential conductance from a linear fit to the current-voltage characteristics for U>10V. Dashed lines are guides to the eye. A kink similar to Fig. 2(c) is observed. (c) Line spectra at $k_\parallel = 0$ for fixed excitation power and increasing bias. Spectra are not normalized and have a constant vertical offset. Due to the pulsed excitation scheme, emission from both polariton and photon lasers can be seen.

To exclude the contribution of a photo-induced voltage due to illumination of the Schottky-contacts, we have recorded current-voltage characteristics at different excitation powers that are plotted in Fig. 3(a). To allow for a better comparison of the curves for different powers, the current was normalized to the total number of excited carriers at each excitation power. For the normalization, we assume 30% sample reflectance in the Bragg minimum used for excitation and 15% absorption in the 12 quantum wells. All curves show a diode-like behavior due to the Schottky-contacts at the metal-semiconductor interface. For $P < 20$mW, the normalized current is similar for all excitation powers. For large excitation powers, there is a clear trend of decreasing normalized current with increasing power. A linear fit to the current-voltage characteristics for U>10V yields the differential conductance as a function of power which is plotted in Fig. 3(b). Here, we again observe a kink at around 20mW like in Fig. 2(c) as expected since differential conductivity and photocurrent are both measures of the same quantity $n_{eh}$ if constant carrier mobility is assumed. This measurement confirms the sudden increase of free carrier decay at the second threshold which is most likely due to stimulated emission of photons.

The effect of the in-plane electric field on PL is illustrated in Fig. 3(c) which shows line spectra at $k_\parallel = 0$ for increasing bias at a constant excitation power of 21.4mW when polariton and photon lasers are comparably bright in the open circuit configuration. Note that spectra are not normalized and are stacked with a constant offset. With increasing bias, emission from the polariton laser continuously decreases resulting in the loss of polariton lasing at around 10V. The decreasing emission from the polariton ground state is due to field ionization of excitons[18,22] which reduces the density of the exciton reservoir feeding the ground state. Emission from the photon mode is unaffected up to 14V and then also quenched at higher voltages as electrons and holes are separated by the in-plane field.

In conclusion, we have combined PL and lateral photocurrent measurements to investigate the power dependence of the free carrier density across two thresholds of the luminescence signal of a quantum well microcavity. We find a notable kink in the photocurrent at the second threshold while the first threshold does not lead to a nonlinearity in the power dependence of the photocurrent. This gives evidence on the absence of free carrier pinning at the polariton lasing threshold as opposed to the second threshold which is identified as a photon lasing transition supported by a fermionic reservoir. Finally, we believe that our technology platform lays ground for more advanced experiments combining PL and electrical measurements. For instance, the fabrication technique introduced here could be used to investigate two-dimensional electron gases coupled to exciton-polaritons where polariton mediated superconductivity is predicted to occur[26].

We thank A.V. Kavokin for helpful discussions and M. Wagenbrenner for assistance during sample growth. This work was supported by the State of Bavaria.